# INFERENCE FOR THE DARK ENERGY EQUATION OF STATE USING TYPE IA SUPERNOVA DATA[1]

By Christopher R. Genovese, Peter Freeman, Larry Wasserman, Robert C. Nichol and Christopher Miller

*Carnegie Mellon University, Carnegie Mellon University, Carnegie Mellon University, University of Portsmouth and CTIO/NOAO*

The surprising discovery of an accelerating universe led cosmologists to posit the existence of "dark energy"—a mysterious energy field that permeates the universe. Understanding dark energy has become the central problem of modern cosmology. After describing the scientific background in depth, we formulate the task as a nonlinear inverse problem that expresses the comoving distance function in terms of the dark-energy equation of state. We present two classes of methods for making sharp statistical inferences about the equation of state from observations of Type Ia Supernovae (SNe). First, we derive a technique for testing hypotheses about the equation of state that requires no assumptions about its form and can distinguish among competing theories. Second, we present a framework for computing parametric and nonparametric estimators of the equation of state, with an associated assessment of uncertainty. Using our approach, we evaluate the strength of statistical evidence for various competing models of dark energy. Consistent with current studies, we find that with the available Type Ia SNe data, it is not possible to distinguish statistically among popular dark-energy models, and that, in particular, there is no support in the data for rejecting a cosmological constant. With much more supernova data likely to be available in coming years (e.g., from the DOE/NASA Joint Dark Energy Mission), we address the more interesting question of whether *future* data sets will have sufficient resolution to distinguish among competing theories.

**1. An accelerating universe.** Current models of the universe posit the existence of a ubiquitous energy field of unknown composition that comprises

Received June 2008; revised December 2008.
[1]Supported ir part by NSF Grant DMS-08-06009.

about 73% of all mass-energy and yet that can only be detected through subtle effects. Cosmologists have dubbed this mysterious field *dark energy*, and over the past decade, it has become an accepted part of the standard cosmology and a focus of observational efforts. More than that: it is fair to say that understanding dark energy has become the central problem in modern cosmology.

In the remainder of this section we explain the scientific background to the problem and describe the quantities that are used in what follows. In Section 2 we discuss several techniques for making inferences about dark energy and examine current results. In Section 3 we describe the data we use, which are measurements of a particular type of exploding star. In Section 4 we formulate the statistical problem as a nonlinear inverse problem

$$(1) \qquad Y_i = T(w)(z_i) + \varepsilon_i, \qquad i = 1, \ldots, n,$$

where the $Y_i$s are scalar observables, $w$ is an unknown function, called the dark energy equation of state, the forward operator $T$ is nonlinear and depends on two unknown parameters, and the $\varepsilon_i$'s are heteroskedastic, uncorrelated noise with known variances. In Section 5 we use features of the forward operator $T$ to construct hypothesis tests that can distinguish among competing cosmological models with minimal assumptions about $w$. In Section 6 we present a framework for computing parametric or nonparametric estimators of $w$ based on equation (1) and compute resampling-based error bars on the estimates. In Section 7 we apply these methods to current data in an attempt to distinguish among competing cosmological models. Finally, in Section 8 we look to the future, where planned observations, both space- and ground-based, will produce much larger data sets that can benefit more fully from these techniques.

The story of how an unobserved and unexplained source of energy came to be quickly and widely accepted as a dominant component of the universe is an essential prelude to discussing the inference problem that is the subject of this paper. To tell this story properly, we need to start earlier, with one of the observational foundations on which all current cosmological models rest: the universe is expanding.

1.1. *Hubble's law and the distance-redshift relation.* In 1929 Edwin Hubble observed a sample of nearby galaxies and studied two quantities for each. The first was the galaxy's redshift.[2] Light from an object that is emitted at one wavelength and observed at a higher (lower) wavelength is said to be redshifted (blueshifted). Astronomers quantify this using a dimensionless parameter $z$, called the *redshift* and given by

$$(2) \qquad z = \frac{\lambda_{\text{obs}} - \lambda_{\text{emit}}}{\lambda_{\text{emit}}},$$

---

[2]Vesto Slipher had first measured these redshifts a decade earlier [Slipher (1917)].

where $\lambda_{\text{emit}}$ is the wavelength of the light measured in the reference frame of the object when the light is emitted and $\lambda_{\text{obs}}$ is the wavelength measured in the reference frame of an observer at some later time. When $z > 0$, the observed light is redshifted relative to the emitted light; when $z < 0$, the observed light is blueshifted relative to the emitted light. The light from an object such as a galaxy contains a mixture of different wavelengths at different intensities, called the spectrum of the object, and redshifting (or blueshifting) causes the entire spectrum to be displaced by a common factor. Because certain features in a galaxy's spectrum, such as emission or absorption lines, have known wavelengths and identifiable patterns, the redshift $z$ can be determined very accurately from observations. Several physical phenomena can cause the redshift to be nonzero, but Hubble interpreted the redshifts as Doppler shifts caused by the relative motion of the galaxies with respect to Earth,[3] where $z > 0$ corresponds to an object moving away from us and $z < 0$ corresponds to an object moving toward us. (Think of an ambulance's siren as it passes you at high speed; the pitch is higher as it approaches and lower as it recedes.)

The second quantity Hubble studied for each galaxy was its distance from us. This he measured himself. Determining accurate distances to faint and far away objects is a challenging, fundamental problem of observational astronomy. The ladder of measurement methods that astronomers have devised can produce reasonably accurate distance estimates in overlapping ranges from nearby stars out to very distant galaxies. Many of these methods rely on having a *standard candle*, a class of celestial objects whose intrinsic brightness is known. If we know how bright an object really is (called its absolute magnitude) and how bright it appears to us (called its apparent magnitude), we can determine its distance from us because the intensity of a light source decays as the inverse square of distance. To determine the distances to the galaxies he observed, Hubble used as a standard candle a class of stars, called Cepheid variables, that fluctuate in brightness with a period that is a function of their average luminosities.

Hubble combined the two measurements—redshift and distance—to produce a plot much like that in Figure 1 [Hubble (1929)]. This shows a strong linear relationship

$$z = \frac{H_0}{c} d, \tag{3}$$

where $c$ is the speed of light and $H_0$ is a fundamental parameter called the Hubble constant. (Note that the linear form of this relationship is in fact an approximation that only holds for small redshifts and distances.) Hubble thus found that galaxies in all directions have redshifts that tend to increase in proportion to their distances, and by interpreting redshifts as Doppler shifts, that galaxies are moving away from us at a velocity $v$ proportional to distance:

$$v = H_0 d. \tag{4}$$

---

[3]Slipher made a similar interpretation: that the "spiral nebulae" he observed were moving away from the Earth. But the claim that such nebulae lay outside the Milky Way was still a matter of intense debate at the time. It was Hubble who settled that argument.

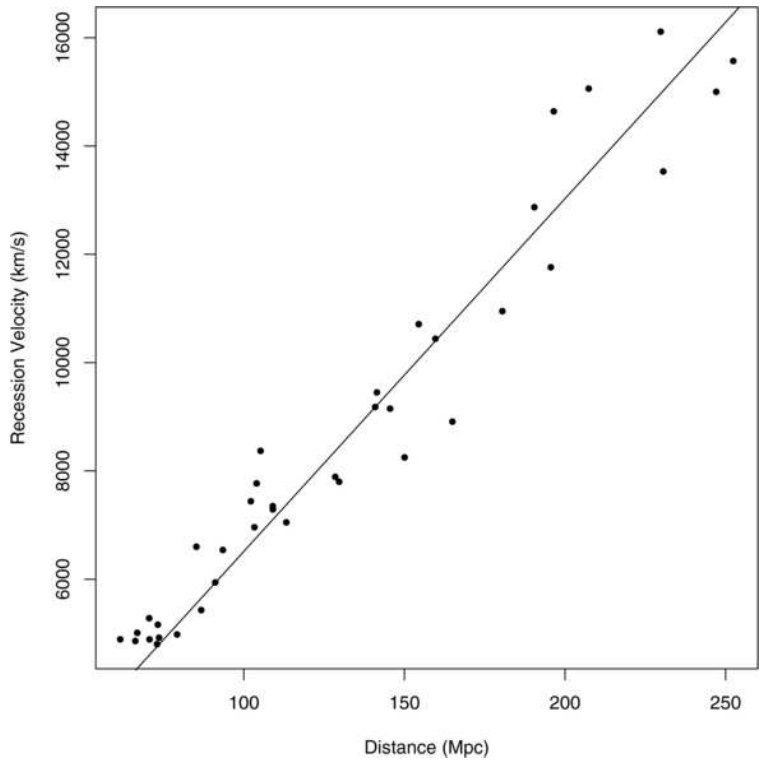

FIG. 1. *A plot analogous to that published by Edwin Hubble but produced with the current data described in Section [3](). The vertical axis shows redshift and the horizontal axis gives a measure of distance. The regression line through the origin highlights the linear relationship in ([3]).*

This relationship is now known as Hubble's law. A natural interpretation of ([4]) would be that Earth lies at the center of an extra-galactic explosion, but this would violate the principle that humans are not "privileged observers" of the universe [Peacock [(1999)]]. Hubble was left to conclude that every galaxy was moving away from every other and, thus, that the universe itself is expanding.

If the universe is expanding, then extrapolating backward in time, the universe was smaller in the past, and thus more dense (with the same total mass-energy packed into a smaller volume), and thus hotter. Taken back far enough, this would predict a point of infinite density and temperature—a singularity known as the "Big Bang."[4] However, basic physical assumptions break down within about $10^{-43}$ seconds of that singularity, so there is currently little known about the singularity itself. What we do know is that the universe began small, dense, and very hot and that it cooled as it expanded.

Extrapolating forward in time, cosmologists could then predict three possible fates for the universe, depending on whether the gravitational pull of all the universe's mass-energy is sufficient to overcome the energy of the Big Bang that is driving the universe's expansion. In an *open* universe, the total mass-energy is insufficient, and the universe will expand forever. In a *closed* universe, the total mass-energy is sufficient to halt the expansion, and the universe will collapse in on itself in a "Big Crunch." On the boundary between these two cases is a flat universe, where gravity and the expansion exactly balance.[5] The search to distinguish among the three cases led to more questions than answers (mostly due to the lack of good standard candles). A leading model for the universe's early expansion [Guth (1981)] engendered a theoretical bias favoring a flat universe; however, the mass and energy needed to slow the expansion could not be found. In fact, studies of galaxies and galaxy clusters suggested that matter accounts for only a fraction of the energy density required for flatness. And observations of the Cosmic Microwave Background (CMB)—relic radiation from several hundred thousand years after the Big Bang when the universe became transparent to light—ruled out a flat universe comprised entirely of matter [Smoot et al. (1992)]. Yet several possibilities remained. Many astronomers thought an open, matter-dominated universe the most likely explanation for the observations. Others found the theoretical arguments compelling and favored a flat universe with either a new form of energy or a new theory of gravity to explain the missing energy density.

Theoretical cosmologists had anticipated an expanding universe several years before Hubble analyzed his data. Alexander Friedmann and George Lemaître independently used Einstein's General Relativity to derive theoretical models that describe the dynamics and geometry of an expanding universe. These evolved into the standard model of relativistic cosmology, called the Friedmann–Robert–Walker model [Peacock (1999)]. Friedmann introduced a dimensionless function called the *scale factor*, $a(t)$, that gives the ratio of the size of any region at time $t$ to the size of that region at the current time $t_0$ [from which it follows that $a(t_0) = 1$]. (Cosmologists often use 0 subscripts for quantities at the current age of the universe.) The function $a(t)$ describes the universe's entire expansion history. Friedmann derived the following equation that describes how the scale factor evolves:

$$\left(\frac{\dot{a}(t)}{a(t)}\right)^2 = \frac{8\pi G}{3}\rho(t) - \frac{kc^2}{a^2(t)}, \tag{5}$$

where $\dot{a}(t)$ is the time derivative of $a(t)$, $G$ is Newton's gravitational constant, $\rho(t)$ is the total energy density at time $t$, and $k$ is a parameter describing the curvature of spacetime, which takes values $-1$, $0$, or $1$ depending on whether the universe is open, flat, or closed.

---

[4]Strictly speaking, the term Big Bang to refers to the general notion that the universe expanded from an early state that was enormously hotter and denser than the present universe. But in common parlance, it is also used to refer to the initial event, as in "years after the Big Bang."

[5] An analogy to an open universe is a rocket that has enough energy to escape Earth's gravity and fly out into space. An analogy to a closed universe is a rocket that cannot escape Earth's gravity and falls back to the ground (or into orbit). An analogy to a flat universe is a rocket with just enough energy to escape to infinity but which would not escape if it had any less.

The Friedmann equation (5) connects the universe's geometry (e.g., size, curvature) to its content (e.g., energy density). An observer at time $t$ who replicated Hubble's study would find a linear relationship between recession velocity and distance to nearby galaxies with slope

$$H(t) = \frac{\dot{a}(t)}{a(t)}. \tag{6}$$

This Hubble parameter $H(t)$ describes the (local) relative expansion rate and is constant for all observers at a particular time; at the current time, we get the Hubble constant $H_0 = H(t_0)$. In a flat universe—when $k = 0$ in equation (5)—the total energy density is determined by $H$ and is called the critical density $\rho_{\text{crit}}$:

$$\rho_{\text{crit}} = \frac{3H^2}{8\pi G}. \tag{7}$$

If $\rho > \rho_{\text{crit}}$, the universe is closed; if $\rho < \rho_{\text{crit}}$, the universe is open; if $\rho = \rho_{\text{crit}}$, the universe is flat. Cosmologists refer to these cases in terms of the scaled density parameter

$$\Omega \equiv \frac{\rho}{\rho_{\text{crit}}}, \tag{8}$$

giving $\Omega > 1$, $\Omega < 1$, and $\Omega = 1$ for closed, open, and flat respectively. At times, it is also useful to decompose the total energy density as a sum of contributions from various sources (e.g., $\rho = \rho_{\text{matter}} + \rho_{\text{radiation}} + \cdots$), and in such cases $\Omega$ has an analogous decomposition. For instance, $\Omega_m = \rho_{\text{matter}}/\rho_{\text{crit}}$ gives the fractional contribution of matter to the critical density.

Relating these theoretical expressions to Hubble's observation, the redshift $z$ can also be expressed in terms of $a$. As the universe expands, light waves traveling through it are stretched out by the expansion increasing the wavelength by the factor by which the universe has expanded between the emission and observation times. That is, light emitted at time $t_{\text{emit}}$ with wavelength $\lambda_{\text{emit}}$ and observed at time $t_{\text{obs}}$ has wavelength $\lambda_{\text{obs}} = \lambda_{\text{emit}}(1+z)$, where the redshift $z$ is given by

$$(9) \qquad z = \frac{a(t_{\text{obs}})}{a(t_{\text{emit}})} - 1.$$

For example, an object observed at $z = 1$ emitted its light when the universe was half its present size. The most distant objects yet observed have $z \approx 6$, when the universe was one seventh its current size. This reveals the redshift to be *cosmological* in origin, the result of the universe's expansion. Because expansion implies $\dot{a}(t) > 0$, redshift can be viewed as an index of time, so it is common for cosmologists to parameterize time-dependent functions interchangeably by time or redshift, for example $H(t)$ or $H(z)$.

Redshift is a measure of distance as well. Hubble's relation (4) defines a frame of reference in which an observer at rest in that frame sees galaxy recession velocities proportional to distance in all directions. Such an observer is said to be *comoving* with respect to the Hubble expansion. In contrast, an observer moving relative to this reference frame would see systematically higher recession velocities behind than in front. The distance measured with a tape-measure between two comoving observers at time $t$ has the form $d(t) = a(t)cr$, where $r$ is called the *comoving distance* between the two observers, which we express in units of time.[6] Each thus sees the other receding at velocity $v(t)$ where, because $r$ does not change with time,

$$(10) \qquad v(t) = \dot{d}(t) = \dot{a}(t)cr = \frac{\dot{a}(t)}{a(t)}d(t) = H(t)\,d(t).$$

This gives the distance-velocity relation, or Hubble's law. But recession velocities are not observable, redshifts are. The comoving distance between an observer and an object at redshift $z$ is derived by computing tape-measure distances between nearby events along the line of sight to the object and adjusting each such distance for the corresponding expansion of the universe. This gives

$$(11) \qquad r(z) = \int_0^z \frac{dz'}{H(z')}.$$

Through equation (11), Friedmann's model ties the distance-redshift relation to the universe's geometry. By carefully measuring objects' redshifts and distances, it is possible to estimate the distance-redshift relation and, in turn, the universe's expansion history, its eventual fate, and a variety of fundamental cosmological parameters. This was the basic task of observational cosmology for many years.

---

[6] Various equally valid conventions for the units of comoving distance are used in the literature, and the choice does not change the results. We express $r$ in units of $H_0^{-1}$, or time. Note that in certain units commonly used by cosmologists, time and distance have the same dimensions because the speed of light provides an absolute conversion between them.

But in 1998, cosmologists discovered something surprising.

1.2. *Acceleration and dark energy.* In 1998 two groups of astronomers [Perlmutter et al. (1998); Riess et al. (1998)] estimated the distance-redshift relation (11) using Type Ia supernovae (SNe), a class of exploding stars whose distance can be measured with ∼15% accuracy, much better than for other distant sources. What they found was that $\dot{a}(t)$ is increasing; in other words, the universe is not merely expanding, the expansion is *accelerating*.

The immediate challenge for astrophysicists was verifying that the apparent acceleration is not an artifact of incorrect assumptions or misinterpretation of the data. Since the initial discovery, many more supernovae have been measured and with greater precision [see, e.g., Davis et al. (2007)], and concerns about systematic errors have been allayed, though not eliminated. Also, indirect supporting evidence comes from measurements of the CMB combined with observations of large scale structure in the distribution of galaxies [Boughn and Crittender (2004); Fosalba et al. (2003); Nolta et al. (2004); Scranton et al. (2003)] and other types of data [see, e.g., Frieman et al. (2008) and the references therein]. These observations all support the hypothesis that the universe is both flat and not comprised entirely of matter; in fact, matter constitutes only about one quarter of the critical density. These results also put stronger constraints on cosmological parameters that in turn sharpen the results of supernova studies. Taken together, current data strongly rule out a nonaccelerating model in comparison to a simple accelerating model.

A more fundamental challenge thus becomes explaining the acceleration.[7] If General Relativity accurately describes physics at large scales and if the universe is homogenous and isotropic at large scales as commonly assumed, then an accelerating universe can be explained by a heretofore unknown type of energy acting against the pull of gravity to speed up the expansion. This energy is characterized by its negative pressure. In contrast, in a universe filled with hot gas, which has positive pressure, the energy of the gas adds to its gravitational account, slowing the expansion. With a negative pressure "fluid," the opposite occurs, causing the universe's expansion to accelerate. Because its source and nature are unknown and because we cannot see it directly, this energy field with negative pressure has been called *dark energy*.

---

[7]For more on explanations of an accelerating universe, the reader is referred to Carroll (2003, 2001), on which much of the following discussion in this subsection is based.

An alternative to the existence of dark energy is that General Relativity or the standard cosmological models built on it do not adequately describe the universe at large scales. General relativity has been strongly tested within the solar system and nearby universe, but not on scales roughly the size of the current universe. Active efforts have been made to modify General Relativity to produce an apparent acceleration. The simplest such modifications have been ruled out by other data and theoretical consistency requirements. More viable modifications have been developed but remain controversial.

Both lines of inquiry remain open, but for now, the evidence appears to strongly support both an accelerating universe and the existence of dark energy. This raises the question of what dark energy is. One possible answer had been introduced by Einstein long-ago for a different reason. At the time that Einstein had proposed his theory of General Relativity—fourteen years before Hubble's observations—it was widely believed that the universe is static. But to achieve a solution of his equations that produced a static (though unstable) universe, Einstein needed to introduce a "cosmological constant" to his equations. He had never been enthusiastic about the cosmological constant because it sullied the pure beauty of his equations, and with Hubble's observations and subsequent evidence for an expanding universe, Einstein ruefully withdrew the cosmological constant, allegedly calling it his "greatest blunder" [Gamow (1970)], possibly because he missed the chance to predict an expanding universe. Years later, particle physicists resurrected the idea of a cosmological constant to represent the energy-density contribution from empty space, a so-called vacuum energy. At cosmological scales, their formulation is mathematically equivalent to Einstein's.

A nonzero cosmological constant can explain an accelerating universe because it acts against gravity and because, as the universe expands, there is more space and thus an increased effect. The *cosmological constant model* specifies a constant vacuum energy throughout time. This is a simple model that is consistent with the available data. There are two problems, however, that suggest a more complicated picture. First, measured values of dark energy are smaller than theoretical predictions of vacuum energy (from quantum field theory) by 120 orders of magnitude [Frieman, Turner and Huterer (2008); Weinberg (1989, 2000)]. Second, according to the cosmological constant model, we live in a time when the total energy density of dark energy is of comparable order to the energy density of matter, which, as we will see below, is a surprising coincidence. Moreover, although the cosmological constant model has an appealing simplicity, there is no known physical reason to require the properties of vacuum energy to be constant in time. Two critical questions then are whether the available data can rule out a cosmological constant in favor of dynamic, time-varying, dark energy and, if so, what the data can tell us about how it varies.

Once we move beyond a cosmological constant, there are relatively few a priori constraints on the dark energy. We can get some information, however, from the fact that the universe is accelerating. Let $\rho(t)$ be the universe's total energy density at time $t$. As stated above, we can decompose $\rho(t)$ into a sum of contributions from different sources, including a dark energy contribution $\rho_{\rm DE}$. The Friedmann equation (5) can be written as

$$\dot{a}^2(t) = \frac{8\pi G}{3} a^2(t) \rho(t) - k, \tag{12}$$

and acceleration implies that $\dot{a}^2(t)$ is an increasing function. It follows that $a^2(t)\rho(t)$ must increase as well and that neither matter nor radiation can be responsible for the acceleration. To see the latter, notice that $\rho_{\rm matter} \propto a^{-3}$ because the total mass of matter in any comoving volume element remains constant while the element's volume increases as $a^3$. Similarly, $\rho_{\rm radiation} \propto a^{-4}$ because expansion redshifts light to higher wavelength and thus lower energy by a factor of $1/a$ and increases volume as $a^3$.

The defining feature of a cosmological constant, however, is that its energy density remains constant, so under this model, $\rho_{\rm DE} \propto a^0$ and $\rho_{\rm DE} a^2$ increases. More generally, we require that $\rho_{\rm DE} \propto a^u$ for $u > -2$. This brings the coincidence problem referred to above into relief. The ratio of the matter density to the dark energy density is given by

$$\frac{\rho_{\rm matter}}{\rho_{\rm DE}} \propto a^{-(3+u)}, \tag{13}$$

so matter dominated in the early universe and dark energy will dominate eventually. Currently, the ratio is about 1/3, which puts us at a relatively unusual time in the life of the universe where the two components are nearly balanced. Cosmologists do not like coincidences, and one hope for dynamical models of dark energy is that they will explain the current balance.

1.3. *The equation of state.* One way to quantify the dark energy is through its energy density $\rho_{\rm DE}$. This is an intuitive quantity that plays a direct role in the equations of cosmological models.

Another convenient quantity that describes dark energy is its *equation of state*, typically denoted by $w$. In physics, an equation of state is a formula that relates several macroscopic observables of a system; an example is the ideal gas law, relating pressure, density, and temperature. In cosmology, dark energy must have a "perfect fluid" equation of state, characterized entirely by its energy density $\rho$ and isotropic pressure $p$. The simplest candidate is

$$p_{\rm DE} = w \rho_{\rm DE} c^2, \tag{14}$$

where $w$ is a property of the dark energy field that may vary across cosmic history. With some abuse of terminology, the ratio $w$ is itself called the "equation of state." In general, $w$ is a function, usually parameterized in terms of redshift as $w(z)$, but for the cosmological constant model, $w$ does not depend on $z$.

For the cosmological constant model, $w(z) \equiv -1$. This model describes a smoothly distributed field with constant energy density and negative pressure. A slightly more general model sets $w(z) \equiv w_0$, where $w_0$ is a constant that need not equal $-1$. In this case, $\rho_{\rm DE} \propto a^{-3(1+w_0)}$ and $a(t) \propto t^{2/(3(1+w_0))}$, so the expansion will accelerate (in a dark-energy dominated universe) if $w_0 < -1/3$. If $w_0 = -1$, the dark energy density stays constant with time; if $w_0 > -1$, it decreases; and if $w_0 < -1$, it grows. Cosmologists often restrict the possible energy-momentum tensors with various "energy conditions" implied by candidate (and somewhat speculative) fundamental constraints on solutions to Einstein's equations. A commonly used such condition requires $w \geq -1$, and most cosmological models follow suit [Carroll et al. (2003)].

1.4. *Measuring dark energy: Type Ia Supernovae.* One of the challenges of dark energy is that it has not yet been observed directly and can only be measured through its subtle influence on other phenomena. Astronomers have developed several methods that are sensitive to the expansion history of the universe and thus depend, directly or indirectly, on the dark energy equation of state. Of these, Type Ia SNe currently provide the best available constraint on the equation of state.

Type Ia SNe are thought to occur when a white dwarf star strips mass off an orbiting companion star until it becomes massive enough to explode. A white dwarf is the remnant of a low- to medium-mass star at the end of its life, with nuclear fusion exhausted, but as it acquires mass from its companion it reaches a threshold above which a supernova occurs. At its peak, the supernova is typically brighter than its entire host galaxy, and it decays in brightness over a span of days or weeks. For reasons that are not yet fully understood, Type Ia SNe all have a similar peak luminosity, raising the hope that they might serve as standard candles. In fact, Type Ia SNe are close to but not quite standard candles because there remains substantial scatter in peak luminosity among nearby SNe. But these SNe also exhibit a strong empirical correlation between peak luminosity and the time it takes them to decrease in luminosity [Phillips (1993)]. Less luminous SNe decay more rapidly, while more luminous SNe decay more slowly. A one-parameter fit reduces scatter in peak luminosity significantly. Taken together, these features make Type Ia SNe valuable cosmological probes: they are bright enough to be detected at great distances and act as "standardizable" candles for distance determination. Potential systematic errors are thought to be smaller than current statistical uncertainties, with the main sources being (i) possible intrinsic differences between Type Ia SNe at low and high redshift and (ii) uncertainty in the extinction/reddening of light caused by dust [Wood-Vasey et al. (2007)]. Observations of Type Ia SNe at different redshifts can thus be used to estimate the distance-redshift relation, and in turn the dark energy equation of state. These are the data we consider in this paper.

**2. Inference for dark energy.** Under mild assumptions,[8] we can express the dark energy pressure $p_{\rm DE}$ and density $\rho_{\rm DE}$ in terms of the co-moving distance $r$ as follows:

$$p_{\rm DE}(z) = -\rho_{\rm crit}\left(\frac{1}{H_0 r'(z)}\right)^2\left(1 + (1+z)\frac{2r''(z)}{3r'(z)}\right), \tag{15}$$

and

$$\rho_{\rm DE}(z) = \rho_{\rm crit}\left[\left(\frac{1}{H_0 r'(z)}\right)^2 - \Omega_m(1+z)^3\right], \tag{16}$$

where $\Omega_m$ is the fractional contribution of matter to $\rho_{\rm crit}$, and $'$ denotes differentiation of $r$ with respect to $z$ [Huterer and Turner (1999)]. Taking the ratio of these functions yields the so-called "reconstruction" equation for the equation of state $w(z)$:

$$w(z) = \frac{H_0^2 \Omega_m (1+z)^3 + (2/3)(1+z)r''(z)/(r'(z))^3}{H_0^2 \Omega_m (1+z)^3 - 1/(r'(z))^2} - 1. \tag{17}$$

A variety of important cosmological models can be expressed in terms of $w$, including the cosmological constant ($w(z) \equiv -1$): topological defect models [frustrated cosmic strings $w(z) \equiv -1/3$ or domain walls $w(z) \equiv -2/3$; Bucher and Spergel (1999)]; various quintessence models [freezing $w'(z) > 0$, Caldwell and Linder (2005), and thawing $w'(z) < 0$, Zlatev, Wang and Steinhardt (1999) and Steinhardt, Wang and Zlatev (1999)]; and even models which allow $w(0) < -1$, such as Cardassian models [Freese and Lewis (2002)] and phantom dark energy [Caldwell, Kamionkowski and Weinberg (2003)].

Several critical questions can be directly addressed with Type Ia SNe data:

1. Are the data consistent with the cosmological constant model?
2. If not, do the data require that the dark energy equation of state varies with time, and if so, how well can we estimate $w$?
3. Do the data rule out any competing theoretical models?

---

[8]In this paper we follow current standard practice and assume a homogeneous, isotropic, and spatially flat universe where matter is nonrelativistic and where gravity is described by General Relativity with the Friedmann–Robertson–Walker metric. Specifically, we do not examine an alternative to the dark energy hypothesis, modified gravity, for which some of these assumptions do not hold; see, for example, Huterer and Linder (2007). For the remainder of this paper, we also use units where the speed of light $c = 1$.

A finding that $w \neq -1$ or, more generally, that $w$ is not constant would rule out the simplest explanation for dark energy, vacuum energy, and would point the way to fundamental new physics. Eliminating some competing theoretical models or producing a sharp estimate of $w$ would strongly constrain the theoretical explanations for dark energy.

The results of studies to date, combining several types of observations, have developed a robust consensus around several findings [Frieman, Turner and Huterer (2008)]. First, there is strong evidence that the universe is accelerating. Second, under the current theoretical framework, there is strong evidence that dark energy exists, with a critical density $\Omega_{\rm DE} \approx 0.76$. And finally, the best available estimates suggest that $w \approx -1$; thus, current data are well fit by the cosmological constant model.

Attention is therefore focused on future Type Ia SNe data sets, which promise to be orders of magnitude larger and, consequently, to provide tighter constraints on $w$. For instance, two particular observatories will begin to collect rich samples of Type Ia SNe during the next decade. The Large Synoptic Survey Telescope (LSST)[9] is a ground-based instrument that will scan the entire sky every few nights and is expected to detect hundreds of thousands of Type Ia SNe per year, while the DOE/NASA Joint Dark Energy Mission (JDEM)[10] is a space-based instrument that will observe thousands of Type Ia SNe, many at relatively high redshift. In addition, additional observations of other indirect probes of dark energy may provide strong constraints on $w$ that complement those produced by supernova data. With so much data, it is expected that the dark energy equation of state will be well determined. (Indeed, systematic errors may come to dominate statistical uncertainties in the inference problem.) But whether this is sufficient depends on the structure of $w$ and on the subtlety of the features that must be determined to distinguish a theoretical explanation for dark energy. Statistical techniques that can efficiently capture complex structure with a minimum of extraneous assumptions will be needed.

Many approaches to estimating $w$ have been used in the literature, but there are three main threads. In the first thread, one assumes that $w$ lies in a specific parametric family $w(z;\theta)$, maps this family through the forward operator in equation (1) to a parametric family for the comoving distance $r(z;\theta)$, and uses maximum likelihood or a comparable criterion to estimate $\theta$. Many authors utilize this approach, most recently including, for example [Barboza and Alcaniz (2008) and Liu et al. (2008)]. Though many parameterizations have been explored, the most common include $w$ constant, $w = w_0$, and $w$ linear in the scale parameter, $w(z) = w_0 + w_1 z/(1+z)$ [the so-called

---

[9]http://www.lsst.org.
[10]http://jdem.gsfc.nasa.gov.

CPL parameterization; Chevallier and Polarski (2001); Linder (2003)]. The number of free parameters is generally limited to two, with arguments made that the data will not constrain additional parameters [Linder and Huterer (2005)]. A variant of this approach is to fit a piecewise constant model to $w$ and use the fitted covariance matrix to rotate into a basis in which the coefficient estimates are independent [Huterer and Cooray (2005)]. In the second thread, one assumes that $r$ lies in some nonparametric class; estimates $r$ and its derivatives by nonparametric smoothing, and uses equation (17) to estimate $w$ [e.g., Daly et al. (2008)]. And in the third thread, one derives the forward operator in terms of the energy density $\rho_{\rm DE}$ instead of $w$ and applies one of the foregoing procedures to estimate $w$ (e.g., Wang and Mukherjee (2004)]. There have also been efforts to eschew the equation of state representation for dark energy and directly estimate kinematic parameters (such as the deceleration function $q$, see Section 4) from which the expansion history can be derived [e.g., Shapiro and Turner (2006)]. Finally, there have been several papers [e.g., Huterer and Starkman (2003), Saini et al. (2004)] that consider using the data to select the parametric model for $w$; we discuss these further in Section 6.

This paper makes two contributions to this line of work. First, we introduce a new technique for testing hypotheses about $w$ that does not require a specific parameterization of $w$. The technique is based on combining shape constraints on $r$, features of the functions in the null hypothesis, and any desired cosmological assumptions. As we show, this technique can be used to distinguish among currently competing models. Second, we develop a framework for nonparametric estimation of $w$ with corresponding assessment of uncertainty. Given a sequence of parametric models for $w$ of increasing dimension, we use the forward operator $T(\cdot)$ to convert it to a sequence of models for $r$ and use the data to select among them. We also show how to construct a representation of $w$ that gives good performance in the forward mapping approach.

Both of these methods take a fully nonparametric approach because, despite many ideas and some theoretical guidance, little is *known* about the function $w$. Although for current data low-dimensional models for $w$ appear sufficient, the same may not be true with future data sets, which will have the precision to detect subtle structure. Accounting for model uncertainty in a rigorous and efficient way is one of the main values added by this work. Such uncertainty is not accounted for by the "figures of merit" proposed by cosmologists, for instance, in Albrecht et al. (2006), Sarkar et al. (2008), and Wang (2008), but it should be. For estimation within a single parametric model, we effectively use a maximum likelihood approach, so our inferences will have a precision comparable to the best methods in current use. But given that the structure of $w$ is unknown and largely unconstrained a priori, some of that precision must be sacrificed to capture that structure at the proper level of complexity. The methods we propose here also satisfy several other needs that cosmologists have. They can straightforwardly be used in combination SNe data with other types of data that probe dark energy. They allow great flexibility in the models and parameterizations used for $w$ and in the cosmological assumptions that can be imposed. They provide an assessment of uncertainty on the inferences. And they are computationally efficient enough to handle the forthcoming large data sets.

**3. Data.** Stellar magnitude is a logarithmic scale for the brightness of astronomical objects, defined so that dimmer objects have larger values. The *apparent magnitude* of an object describes how bright the object appears from Earth: $m = 2.5 \log_{10}(f/f_0)$, where $f$ is the flux of light produced by the object that is received by a detector on Earth (in some specified range of wavelengths) and $f_0$ is the corresponding flux for a reference object. The *absolute magnitude* of an object describes how bright the object appears from a fixed reference distance (10 parsecs), and thus is related to the apparent magnitude by $M = m - 5\log_{10}(d_L/10)$, where $d_L$ is a distance to the object measured in parsecs. [Specifically, $d_L$ is called the luminosity distance, which is one of several metrics used by astronomers; cf. Hogg (2000), and which differs from comoving distance by a factor of $1+z$.] The difference between them, $\mu = m - M = 5\log_{10}(d_L/10)$, is called the *distance modulus*, a logarithmic measure of the distance to the object. For Type Ia SNe, apparent magnitudes are observed directly, and absolute magnitudes are determined from the observed luminosities over SNe lifetimes by fitting these observed "light curves" to a template and estimating their peak luminosities [Wood-Vasey et al. (2007)].

We analyze data for 192 SNe Ia from Davis et al. (2007) [see also Riess et al. (2006) and Wood-Vasey et al. (2007)]. The data includes, for each supernova: (i) redshift $z$, (ii) distance modulus $\mu$, and (iii) standard error $\tau = \sqrt{\tau_\mu^2 + \tau_v^2}$ for the distance modulus, where $\tau_\mu$ is the intrinsic uncertainty in the distance modulus and $\tau_v$ is an estimate of error induced by the supernova's peculiar velocity relative to a local standard of rest. (We ignore the uncertainties of the redshift estimates, which are generally less than 1%.) In the Supplementary Material [Genovese et al. (2009)], we also identify two supernovae whose data exhibit nontrivial influence in the analyses, but we include them because they could not be disqualified on concrete grounds.

Let $U_i$ and $z_i$ denote the observed distance modulus and redshift, respectively, for the $i$th supernova, $i = 1, \ldots, n$, where $n = 192$ and where, contrary to common astronomical practice, we have ordered that data by the redshift rather than date of supernova. We model $U_i$ as Gaussian with mean $\mu(z_i)$, that is,

$$U_i = \mu(z_i) + \tau_i \varepsilon_i, \qquad (18)$$

where the $\varepsilon_i$s are assumed independent, mean zero, Gaussian noise terms with unit variance and the $\tau_i$s are the given standard errors of the distance moduli measurements.

We express the data in terms of comoving distance (assuming a flat universe) by transforming as follows:

$$\text{(19)} \qquad \frac{1}{c(1+z_i)} 10^{(U_i-25)/5} = r(z_i) \cdot 10^{(\tau_i/5)\varepsilon_i},$$

where $c$ is the speed of light. Thus, letting $Y_i$ denote the $\log_{10}$ of the left-hand side of (19), we have

$$\text{(20)} \qquad Y_i = \log_{10} r(z_i) + \sigma_i \varepsilon_i, \quad i = 1, \ldots, n,$$

where $\sigma_i = \tau_i/5$. Figure 2 shows $10^Y$ plotted against $z$ with associated error bars; such a plot is called a "Hubble diagram." We thus call $r$ "observable" because it can be directly estimated from the observed data.

**4. Nonlinear inverse problem formulation.** We distinguish two uses of the word *model* in this paper. A cosmological model for dark energy is a set of assumptions about the underlying physics that gives rise to a particular form of the equation of state. A statistical model for $w$ is a family of probability distributions for the data indexed (at least) by a parameterization of $w$, possibly infinite dimensional.

A particular cosmological model can be analyzed under a specific statistical model, but the scope of the inferences is limited by the viability and flexibility of the assumptions made. We consider statistical models whose stochastic component is specified by equation (20); each such model is then determined by the parameters $H_0$ and $\Omega_m$ and a representation of $w$.

We now re-express the relationship between the comoving distance and the equation of state as an explicit analytic expression mapping $w$ (and the cosmological parameters $H_0$ and $\Omega_m$) to $r$. Equation (11) describes the relationship between comoving distance and redshift. By expanding $H(z)$ as in Huterer and Turner (2001), one can derive the following equation: (we give an alternative derivation in the on-line Supplementary Material)

$$\text{(21)} \qquad r(z) = H_0^{-1} \int_0^z ds [\Omega_m (1+s)^3 + (1-\Omega_m)(1+s)^3 \\ \times e^{-3 \int_0^s -w(u)/(1+u)\, du}]^{-1/2}.$$

We can thus write $r$ as the image of an operator acting on $w$; specifically, we can write $r = 10^{T(w;H_0,\Omega_m)}$, where the nonlinear operator $T(w;H_0,\Omega_m)$ is defined by $\log_{10}$ of the right-hand side of equation (21). Equation (20) thus describes a nonlinear inverse problem

$$\text{(22)} \qquad Y_i = T(w; H_0, \Omega_m)(z_i) + \sigma_i \varepsilon_i, \quad i = 1, \ldots, n.$$

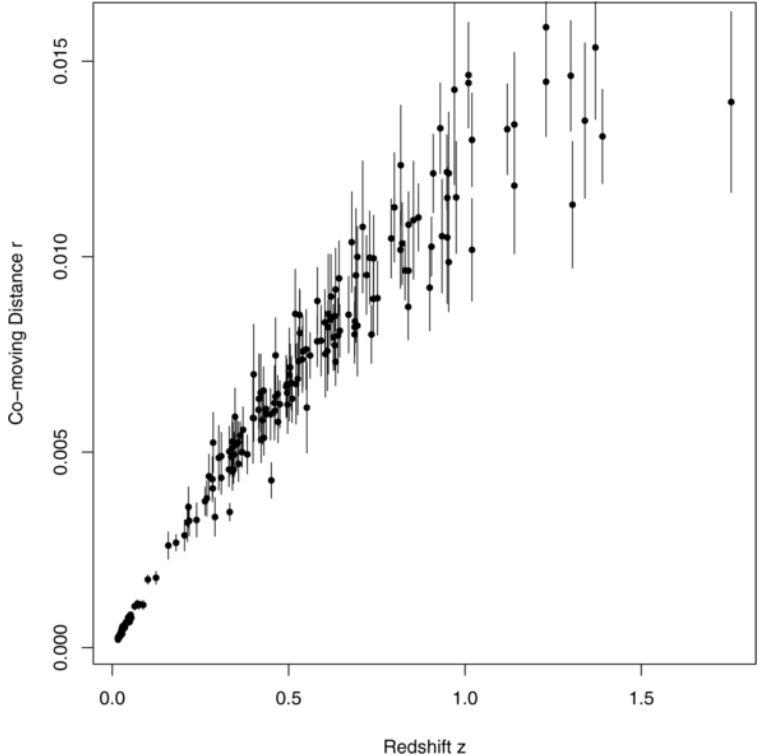

FIG. 2. *The data from equation (20) transformed to plot comoving distance versus redshift. Standard error bars are attached to each point.*

The operator $T$ is complicated and depends on two unknown parameters, but it does have several useful properties, as we will see below. An important use of equation (21) is to translate a model for the unobservable $w$ into the observable $r$. For instance, given any parameterization of $w$, equation (22) determines a likelihood function. We show how to use this for parametric or nonparametric inferences in the following sections.

It is also sometimes helpful to consider the universe's acceleration directly. Cosmologists traditionally express this in terms of the dimensionless *deceleration parameter*[11] $q(z)$, which is defined in terms of the scale factor by

$$(23) \qquad q(t) = -\frac{\ddot{a}(t)a(t)}{\dot{a}^2(t)} = -1 - \frac{\dot{H}(t)}{H^2(t)},$$

---

[11]This was named when it was thought that the universe was decelerating.

and in terms of redshift and comoving distance by

$$(24) \qquad q(z) = -1 - (1+z)\frac{r''(z)}{r'(z)}.$$

Using a similar method to the above, we can solve for $q$ in terms of $w$ and for $r$ in terms of $q$ to obtain

$$(25) \qquad q(z) = \frac{1}{2} + 3w(z)\frac{1 - \Omega_m}{1 - \Omega_m + \Omega_m e^{3\int_0^z -w(s)/(1+s)\,ds}},$$

$$(26) \qquad r(z) = H_0^{-1}\int_0^z ds\, e^{-\int_0^s du(1+q(u))/(1+u)}.$$

These equations are derived in the Supplementary On-line Material.

Equations (21), (25) and (26) have several properties that are valuable for statistical inference. First, note that for any $w$, $r'(z) > 0$, so $r$ is a monotone increasing function of $z$ with $r(0) = 0$. In fact,

$$(27) \qquad \frac{(1+z)^{-3/2}}{H_0} \leq r'(z) \leq \frac{(1+z)^{-3/2}}{\sqrt{H_0^2 \Omega_m}},$$

where the upper bound can be made sharper if $w$ is assumed bounded from below. Second, $r$ is monotone decreasing in $w$ for each fixed value of $H_0$ and $\Omega_m$. Specifically, if $w_1$ and $w_2$ are two candidate equations of state with corresponding comoving distance functions $r_1$ and $r_2$, and if $w_2(z) \geq w_1(z)$ for all $z \geq 0$, then $r_2(z) \leq r_1(z)$ for all $z \geq 0$. Similarly, $q$ is monotone increasing in $w$; $w_2(z) \geq w_1(z)$ for all $z \geq 0$ implies that $q_2(z) \geq q_1(z)$ for all $z \geq 0$. And $r$ is monotone decreasing in $q$; $q_2(z) \geq q_1(z)$ for all $z \geq 0$ implies that $r_2(z) \leq r_1(z)$ for all $z \geq 0$. Third, as shown in the Appendix, for $r$ to be concave it is sufficient that $w(z) \geq -1/(1-\Omega_m)$ for all $z \geq 0$. Under mild smoothness assumptions on $w$, the concavity of $r$ holds more broadly. Fourth, in both equations involving $w$, taking $w \equiv 0$ is equivalent to taking $\Omega_m = 1$.

For any specific parameterization of $w$ and any choice of $\Omega_m$ and $H_0$, it is straightforward to evaluate $r$ numerically. For instance, under a constant $w$ model $w \equiv w_0$, equation (21) reduces to

$$(28) \qquad r(z) = H_0^{-1}\int_0^z ds[\Omega_m(1+s)^3 + (1-\Omega_m)(1+s)^{3(1+w_0)}]^{-1/2}.$$

More generally, expanding $w(z) = -\sum_j \beta_j \psi_j(z)$ in a (not-necessarily orthonormal) basis $\psi_0, \psi_1, \ldots$ yields

$$(29) \qquad r(z) = H_0^{-1}\int_0^z ds[\Omega_m(1+s)^3 + (1-\Omega_m)(1+s)^3 e^{-3\sum_j \beta_j \tilde{\psi}_j(s)}]^{-1/2},$$

where $\tilde{\psi}_j(s) = \int_0^s \psi_j(u)/(1+u)\,du$. Taking the expansion to be finite gives three important special cases:

1. Polynomial in $z$: $\psi_j(z) = z^j$, $j = 0, \ldots, d$, giving

$$r(z) = H_0^{-1} \int_0^z ds [\Omega_m (1+s)^3 \tag{30}$$
$$+ (1 - \Omega_m)(1+s)^{3(1-\alpha_0)} e^{-3\sum_{j=1}^d (-1)^j \alpha_j s^j / j}]^{-1/2},$$

where $\alpha_k = \sum_{j=k}^d (-1)^j \beta_j$ for $k = 0, \ldots, d$.

2. Polynomial in the scale factor $a$: $\psi_j(z) = (1+z)^{-j}$, $j = 0, \ldots, d$, giving

$$r(z) = H_0^{-1} \int_0^z ds [\Omega_m (1+s)^3 \tag{31}$$
$$+ (1 - \Omega_m)(1+s)^{3(1-\beta_0)} e^{3\sum_{j=1}^d \beta_j ((1+z)^{-j} - 1)/j}]^{-1/2}.$$

3. Piecewise constant: $\psi_j(z) = 1_{(s_j, s_{j+1}]}(z)$ for $j = 0, \ldots, K-1$, where $0 = s_0 < s_1 < \cdots < s_K$ are breakpoints for $K$ fixed bins and where $1_{(s_j, s_{j+1}]}(z)$ is 1 if $s_j < z \leq s_{j+1}$ and 0 otherwise. In this case, equation (21) becomes

$$r(z) = H_0^{-1} \int_0^z ds [\Omega_m (1+s)^3 + (1 - \Omega_m)(1+s)^3 e^{-3B(s)}]^{-1/2}, \tag{32}$$

where

$$B(s) = \sum_{j=1}^{J(s)} \beta_j \log\left(\frac{1+s_j}{1+s_{j-1}}\right) + \beta_{J(s)+1} \log\left(\frac{1+s}{1+s_{J(s)}}\right) \tag{33}$$

and where $J(s) = \max\{0 \leq j \leq K : s_j \leq s\}$. Despite the discontinuities in $w$, this expression is a smooth function of the $\beta$ parameters.

Extension to other bases—such as B-splines, orthogonal polynomials, and wavelets—is straightforward.

Combined with equation (20), each of these expressions produces a likelihood for $w$, $H_0$, and $\Omega_m$. Although nonlinear, these likelihoods are well-behaved for optimization purposes, and weighted, nonlinear least-squares is computationally efficient in practice. Good estimates of the coefficients can be obtained for a wide variety of models, which in turn supports both parametric and nonparametric inferences about $w$.

**5. Methods I: hypothesis testing.** As discussed above, one method for distinguishing among models of dark energy is to first estimate the equation of state and use this estimate to test hypotheses about cosmological models [e.g., Huterer and Cooray (2005)]. This approach has several disadvantages, including that the power of the test depends on having a good estimator and that it requires accurate standard errors for the entire function. Moreover, in practice, such tests usually rely for their validity on an assumed parametric model for $w$. It would be desirable to be able to test cosmological models without a preliminary estimator or assumed parameterization, and in this section, we construct a method to do that for certain classes of hypotheses.

The basic idea is that we use the forward operator given in equation (21) to map a set of possible $w$'s to the $r$ domain, and use the data to test the hypothesis there by inverting a nonparametric confidence set for $r$. Two issues arise in such a scheme. First, for general sets in the $w$-domain, it can be difficult to compute their image in the $r$-domain, but we use the properties of the operator discussed in Section 4 to easily compute the mappings for certain classes of hypotheses. Second, performing a sharp nonparametric test (or constructing a small nonparametric confidence set) can be difficult without structural assumptions, but we take advantage of strong shape constraints satisfied by the comoving distance function.

Here, we consider null hypotheses of the following forms:

A. simple equalities for $w$: $w = w_0$,
B. inequalities for $w$: $w_0 \leq w \leq w_1$,
C. inequalities for $w'$: $w'_0 \leq w' \leq w'_1$,
D. inclusion: $w \in V$ for a linear space $V$ of fixed dimension,

and various intersections of these, where $w_0$, $w_1$, $w'_0$, and $w'_1$ denote various fixed functions, not necessarily constant. [We use the inequality $w \leq w_0$ to mean that $w(z) \leq w_0(z)$ for all $z$, and similarly for other inequalities between functions.]

Testing such hypotheses gives direct tests of various cosmological models. The null hypothesis that the cosmological constant model holds, for example, translates to a simple null hypothesis with $w_0 = -1$. Quintessence solutions lead to a variety of constraints on $w$ and $w'$ that can be tested by combining hypotheses that are inequalities for $w$ and for $w'$. For instance, as we show in the Appendix, thawing solutions satisfy

$$(34) \qquad 1 + w \leq \frac{dw}{d\ln a} \leq 3(1+w),$$

and freezing solutions satisfy

$$(35) \qquad 3w(1+w) \leq \frac{dw}{d\ln a} \leq 0.2w(1+w),$$

when $-1 \leq w \lesssim -0.8$ [Caldwell and Linder (2005)], where $a$ is the scale factor. These bounds can be re-expressed for $w$ in the same range as

$$(36) \qquad \frac{1+w(0)}{(1+z)^3} - 1 \leq w(z) \leq \frac{1+w(0)}{1+z} - 1,$$

and

$$(37) \qquad \frac{w(0)}{(1+z)^3 + w(0)((1+z)^3 - 1)} \leq w(z) \leq \frac{w(0)}{(1+z)^{0.2} + w(0)((1+z)^{0.2} - 1)},$$

where $w(0)$ is a free parameter.

The strategy underlying our testing procedure is to use equation (21) to translate hypotheses about $w$ into hypotheses about $r$, making it possible to test any hypothesis that translates into a manageable form. Our test is derived by inverting a $1 - \alpha$ confidence set for $(r(z_1), \ldots, r(z_n))$, as follows:

0. Select a small $0 < \alpha < 1$.
1. Construct a $1-\alpha$ confidence set $\mathcal{C}$ for the unknown vector $(r(z_1), \ldots, r(z_n))$.
2. Construct the set $R_0$ of vectors $(r_0(z_1), \ldots, r_0(z_n))$ where $r_0$ is a comoving distance function produced by an equation of state consistent with the null hypothesis.
3. Reject the null hypothesis if $\mathcal{C} \cap R_0 = \varnothing$.

In practice, the sets in steps 1 and 2 need not be constructed explicitly. For example, $\mathcal{C}$ typically takes the form of bands—an interval at each $z_i$—or a ball centered around a particular vector. And $R_0$ can usually be represented implicitly by an efficient search over the null hypothesis in $w$ (mapped forward by $T$) targeting those $r$'s that lie outside $\mathcal{C}$. In practice, this procedure can be made computationally efficient for a broad range of hypotheses.

One way to define the confidence set $\mathcal{C}$ is the set of vectors for which a standard chi-squared goodness-of-fit test does not reject the null hypothesis. In light of equation (20), the chi-squared goodness-of-fit ball gives a confidence set for $(\log_{10}(r(z_1)), \ldots, \log_{10}(r(z_n)))$, which is easily transformed into a confidence set for $(r(z_1)), \ldots, r(z_n))$. As the number of data grows, however, the chi-squared confidence sets become unduly conservative, reducing the power of the test. So, we also use alternative confidence set procedures that produce smaller confidence sets, giving the test higher power [Baraud (2004), Davies et al. (2007), Ingster and Suslina (2006)].

Suppose that for $i = 1, \ldots, n$, $Y_i = f_i + \sigma_i \varepsilon_i$, where the $\sigma_i$s are known numbers and the $\varepsilon_i$s are independent Gaussian variables. This corresponds to equation (20) with $f_i = \log_{10} r(z_i)$. If the vector $f = (f_1, \ldots, f_n)$ denotes the true but unknown values of the function at the observed points, then a $1 - \alpha$ confidence set $\mathcal{C}$ for $f$ is a random set, constructed from the data, that satisfies

$$P\{\mathcal{C} \ni f\} \geq 1 - \alpha. \tag{38}$$

We want $\mathcal{C}$ to be as small as possible. (Although there are several reasonable definitions of size, we will use the simplest: the radius of a confidence ball in the corresponding norm, the width of confidence bands, and the volume of a more general set. Our methods work as well if the confidence set is constructed to optimize some other criterion.)

There are several ways to construct such confidence sets. One way is to invert a chi-squared goodness-of-fit test for the null hypothesis $f = f^0$, giving $\mathcal{C} = \{f^0 : T^2(f_0) \leq \chi^2_{n,\alpha}/n\}$, where $\chi^2_{n,\alpha}$ is the upper-tail $\alpha$ quantile of the corresponding chi-squared distribution and where $T^2(f_0) = (1/n)\sum_{i=1}^{n}(Y_i - f_i^0)^2/\sigma_i^2$.

The chi-squared confidence set is simple to use, but it has several major drawbacks. The confidence set is relatively large; the radius of the set $\chi_{n,\alpha}/\sqrt{n}$ is $O(1)$ no matter how large $n$ is. The set is constructed from a rough estimator of $f$, namely, the data. The size of the set is independent of the data and thus cannot adjust to evidence of smoothness. And some prior information, such as shape restrictions, is difficult to incorporate in practice.

There are practical confidence set procedures that address all these drawbacks. We consider two: shape-restricted confidence bands from Davies, Kovac and Meise (2007) and adaptive chi-squared confidence sets from [Baraud (2004)]. Both provide finite-sample, honest [Li (1989)] confidence sets that adapt their size based on the data. Both provide substantially smaller confidence sets than the chi-squared ball. Both are computationally practical, though somewhat more work than the naive chi-square confidence set. And both allow us to incorporate prior information about the comoving distance to produce a smaller confidence set. Because the Davies et al. procedure performed better in simulations with current sample sizes and because it requires fewer tuning parameters, we focus on that procedure in this paper. The shape constraints are also well adapted to our prior information about $r$. In principle, the Baraud procedure combined with shape constraints should outperform the Davies et al. procedure, but we will explore this comparison in a future work. More detail about the Baraud procedure is given in the Supplementary On-line Material.

The procedure of Davies et al. generates confidence bands under the assumption that $f$ is monotone and concave. For each $1 \leq i \leq k \leq n$, define the integer interval $I_{ik}$ to be the set of indices $1 \leq j \leq n$ such that $z_i \leq z_j \leq z_k$. Thus, if there are no ties among the redshifts, $I_{ik} = \{i, \ldots, k\}$. Consider the following statistics, modified from Davies, Kovac and Meise (2007), to account for the different standard errors of the measurements:

$$
(39) \qquad T_{ik}(f) = \frac{1}{\sqrt{\#(I_{jk})}} \sum_{j \in I_{ik}} \frac{Y_j - f_j}{\sigma_j},
$$

where $\#(\cdot)$ gives the cardinality of a set. These serve as test statistics for testing whether the residual mean is zero along any index interval. When $f$ is the true vector, the $T_{ik}(f)$s each are mean zero Gaussian variables. Davies, Kovac and Meise (2007) point to a procedure for computing these statistics in $O(n \log n)$ time and offer an approximating subset of size $O(n)$ for large $n$. We use the latter for convenience, but use of the full set did not significantly change the results.

The procedure begins with a confidence set for the $T_{ik}$'s. In Davies, Kovac and Meise (2007), this is a confidence cube with edge length equal to twice the $1-\alpha$ quantile of $\max_{i,k}|T_{ik}(f)|$. The key for constructing the final confidence set is that the initial confidence set has linear boundaries, making consistency with the initial confidence set a computationally tractable constraint. Constraints for concavity and monotonicity are also linear. This gives a confidence set for $f$ consisting of those vectors $g$ such that the $T_{jk}(g)$s lie in the initial confidence set and that $g$ satisfies the imposed shape constraints. We could use this confidence set directly, but it is computationally much simpler with the above types of hypotheses to use confidence bands. We compute the confidence bands are constructed by maximizing and minimizing $f_j$ subject to the vector $f$ lying in the initial confidence set and satisfying the shape restrictions. These require optimization of a linear object function with linear constraints and thus can be solved with two linear programs (i.e., optimization problems with a linear objective and linear constraints) for each $j$.

We modify the Davies, Kovac and Meise (2007) procedure in several ways. First, because monotonicity and concavity are used in the procedure in log space, the confidence bands need not be concave in $r$ space. We adjust for this by optimizing the bands in $r$ space, finding the smallest bands consistent with the shape restrictions there, including the additional constraint that $r(0) = 0$. This involves two additional linear programs for each $z_i$. Second, we can use a smaller initial confidence set for the $T_{ik}$s with some additional computation. Specifically, the distribution of $T_{ik}(f)$'s is a degenerate Normal whose covariance depends only on the collection of $I_{ik}$ intervals. With an eigen-decomposition of this covariance, we replace the hyper-cube of Davies, Kovac and Meise (2007) by a substantially smaller degenerate ellipsoid and get proper coverage. The $O(n)$ sized symmetric eigen-decomposition is expensive for large $n$ but can be parallelized if necessary. This also requires optimizing a linear function subject to linear and ellipsoidal constraints. Such a problem can be reduced to a convex optimization problem called second-order cone programming [Boyd and Vandenberghe (2004)], in which we minimize a linear function $\gamma^T x$ subject to one or more "second-order cone constraints" of the form $\|Ax + b\| \leq c^T x + d$, where $\|\cdot\|$ is the Euclidean norm, $A$ is a matrix, $b$ and $c$ are vectors, and $d$ is a scalar. Many common convex optimization problems can be reduced to this form. We used the MOSEK and CPLEX software[12] to compute the solutions to these problems. Finally, to test whether a linear space of vectors intersects the ellipsoidal constraint set and satisfies the shape constraints, we can minimize the (quadratic) distance to the center of the ellipsoid subject to the linear shape constraints and inclusion in the given space. If the resulting optimum is sufficiently close to the center, the test is affirmative. This is yet another form of convex optimization (called quadratic programming) that can be implemented with the same software.

---

[12]See http://www.mosek.com and http://www.ilog.com/products/cplex/.

In all of the hypotheses we test, there are one to three free parameters that must be varied, usually including $H_0$ and $\Omega_m$. A simple grid search is practical and straightforward in these cases. For any value of the free parameters, we can tell whether the corresponding $r$ lies in the confidence bands by direct comparison. We can also incorporate information from other studies by using confidence sets on $(H_0, \Omega_m)$ to define this subsidiary search. Assuming independent data sets and taking $\alpha' = 1 - \sqrt{1-\alpha}$, if we use a $1 - \alpha'$ confidence set for the cosmological parameters and in our procedure, the resulting test has level $\alpha$ as required.

Our procedure works also under more restrictive assumptions about the form of $w$, with correspondingly sharper results as the assumptions grow stronger. For this, the confidence set $\mathcal{C}$ is constructed using the assumed parameterization. The resulting test will have higher power than the nonparametric test when the assumed parameterization holds. Note, however, that the validity of *any* inferences under a specific parameterization depends strongly on the parameterization being accurate. For instance, we assume here that $w \geq -1$ [Carroll, Hoffman and Trodden (2003)], which implies that $r$ is monotone concave. But the same basic procedure works, with somewhat lower power, when that assumption is dropped.

Step 2 of the procedure depends specifically on the hypothesis being tested. We now derive the sets $R_0$ for null hypotheses of the forms listed above. Let $\mathcal{M}$ denote the set of vectors $(r(z_1), \ldots, r(z_n))$ for functions $r$ that meet the a priori conditions that the comoving distance must satisfy:

A. Under a simple null hypothesis $w = w_0$, equation (21) generates a two-parameter family of functions $r_0$ as $H_0$ and $\Omega_m$ vary; $R_0$ is the set of vectors $(r_0(z_1), \ldots, r_0(z_n))$ for $r_0$ in this family.
B. Under the null hypothesis, $w \geq w_0$, equation (21) shows that, for fixed $H_0$ and $\Omega_m$, $r \leq r_0$, where $r_0$ is produced in (21) by $w = w_0$ for the given value of $H_0$ and $\Omega_m$. Again, varying $H_0$ and $\Omega_m$ produces a two-parameter family of functions $r_0$. $R_0$ is the set of vectors $(r_1, \ldots, r_n) \in \mathcal{M}$ such that $r_1 \geq r_0(z_1), \ldots, r_n \geq r_0(z_n)$ for some $r_0$ in the family. The restriction to $\mathcal{M}$ sharpens the results. It is not strictly necessary, but because equation (21) produces functions in $\mathcal{M}$, it is an improvement that is virtually cost free. The other direction of inequality is handled similarly, using the monotonicity of $r$ in $w$.
C. Null hypotheses of the form $w' \geq w'_0$ can be handled by re-expressing the exponent in equation (21). Integrating by parts and writing $w(s) =$

$w(0) + \int_0^s w'(u)\,du$ yields that

(40) $\int_0^s \frac{w(u)}{1+u}\,du = w(0)\log(1+s) + \int_0^s w'(u)(\log(1+s) - \log(1+u))\,du.$

  The second term in the right-hand side integrand is nonnegative, so $w' \geq w'_0$ implies, for fixed $H_0$, $\Omega_m$, and $w(0)$, that $r \leq r_0$, where $r_0$ is the right-hand side of equation (21) corresponding to $(w(0), w'_0, H_0, \Omega_m)$. Varying $H_0$, $\Omega_m$, and $w(0)$ produces a three-parameter family of functions, and as before, $R_0$ is the set of vectors in $\mathcal{M}$ whose components are at least as big everywhere as some function in this family. Other inequalities in $w'$ are handled similarly.

D. The null hypothesis that $w$ lies in some linear space of functions $V$ is useful primarily to test the goodness of fit of statistical models for $w$. We select an arbitrary basis for $V$ and form a $\dim(V) + 2$ dimensional family of functions corresponding to each $(H_0, \Omega_m)$ and each vector of coefficients in the basis expansion. $R_0$ is the set of vectors produced by these functions evaluated at $z_1, \ldots, z_n$. See equation (29). This case is handled in practice by numerical optimization and thus works best for low to moderate dimensional spaces. It is not necessary to restrict to a linear space, but that is the best behaved case numerically.

Note that this same approach can be used to test hypotheses about the deceleration function $q(z)$. For instance, we may wish to test the null hypothesis that the universe is nonaccelerating. This can be expressed in several ways. First, we can test whether the universe is matter dominated without dark energy, which corresponds to $w = 0$, or equivalently, $\Omega_m = 1$, or $q = 1/2$ from equation (25). We call this the matter dominated hypothesis. In contrast, a nonaccelerating universe corresponds to $q \geq 0$; we call this the strongly nonaccelerating hypothesis. Using the monotonicity properties of equation (26), this matter-dominated hypothesis maps to the null hypothesis $r = H_0^{-1}(2 - 2(1+z)^{-1/2})$ and the pure nonaccelerating hypothesis corresponds to the null hypothesis $r \geq H_0^{-1}\log(1+z)$. Both generate a one-parameter family and corresponding $R_0$. Note two issues with the latter, one-sided hypothesis. By taking $H_0$ large enough, we can make the lower bound as small as possible and the null trivially true, so we need to use a prior confidence set for $H_0$ as described above. Second, the condition $q \geq 0$ allows expansion histories that are strongly at odds with current theory. So, strictly speaking, testing $q \geq 0$ will offer poor power against alternatives we care about. We fix this problem in two ways: combining $q \geq 0$ with (i) a bound on $w$ such as $w \geq -1/3$ to preclude acceleration in a dark-energy dominated universe, and (ii) assume that $q$ exhibits a change point between the matter-dominated case $q = 1/2$ and the dark-energy dominated case $q \geq 0$. Both tighten the bounds substantially, at the expense of an added restriction or free parameter.

**6. Methods II: estimating the equation of state.** Answering the main questions about dark energy that are currently being addressed with supernova data involves testing among competing models. But ultimately we want to estimate the equation of state. As described earlier, this problem has received substantial attention in the literature. In this section we describe a framework for constructing nonparametric estimators of $w$ with an associated assessment of uncertainty.

A nonparametric procedure for estimating the equation of state treats $w$ as an infinite parameter belonging to a specified space of functions. Of course, with only a finite amount of data, any estimator has limited resolution, so we must use the data to determine the complexity of the estimator. This gives rise to the bias-variance trade-off: too high a complexity provides a better apparent fit but gives estimates with high variance; too low a complexity gives estimates with low variance but bias from model misspecification.

There have been several related works in the dark energy literature. Saini, Weller and Bridle (2004) formulate the problem as a Bayesian inference problem where the parameter space includes disjoint spaces of low-dimensional polynomials. Huterer and Starkman (2003) construct an empirical basis via principal components. They correctly note that choosing the number of basis elements in a nonparametric analysis requires that one carefully balance the bias and the variance. We provide a concrete mechanism for choosing the number of elements in practice.

Our approach begins with a collection of models $\mathcal{M}_1, \mathcal{M}_2, \ldots$, where each $\mathcal{M}_k$ is a linear space of functions with dimension $k$. We then select a model that balances bias and variance by minimizing an empirical measure of risk or comparable criterion. We use BIC [Bayesian Information Criterion, also called the Schwarz criterion; Schwarz (1978)], as it is both simple and effective, but other criteria give similar results. All the spaces $\mathcal{M}_k$—including those for small $k$—contain smooth and constant functions, and as $k$ increases, the spaces add detail to capture more complex fluctuations. We define $\mathcal{M}_k$ to be the $k$-dimensional space of cubic B-splines [de Boor (2001)] over the range of the data, with equally spaced knots. The choice of knots could be optimized for even better performance by adapting it to the distribution of redshifts in the data. Note that with this choice, the bases for the $\mathcal{M}_k$ spaces are not nested.

A common alternative choice in nonparametric function estimation is to choose an orthonormal basis to represent $w$ and define $\mathcal{M}_k$ to be the span of the first $k$ terms. In an inverse problem such as this, however, this strategy requires care. Choosing a basis to obtain an efficient estimator requires balancing the information passed by the forward operator and the conciseness of the representation for $w$. For example, the Fisher basis [e.g., Huterer and Cooray (2005)] based on the forward operator in equation (22) has most of its variation at low to moderate redshift where $w$ is relatively more informative in the data. But as shown in Figure 3, the basis elements decay quickly to zero. To fit a simple smooth function, even a constant, requires a large number of basis functions, leading to high variance estimators. This is an especially important issue when considering variations around the cosmological constant model.

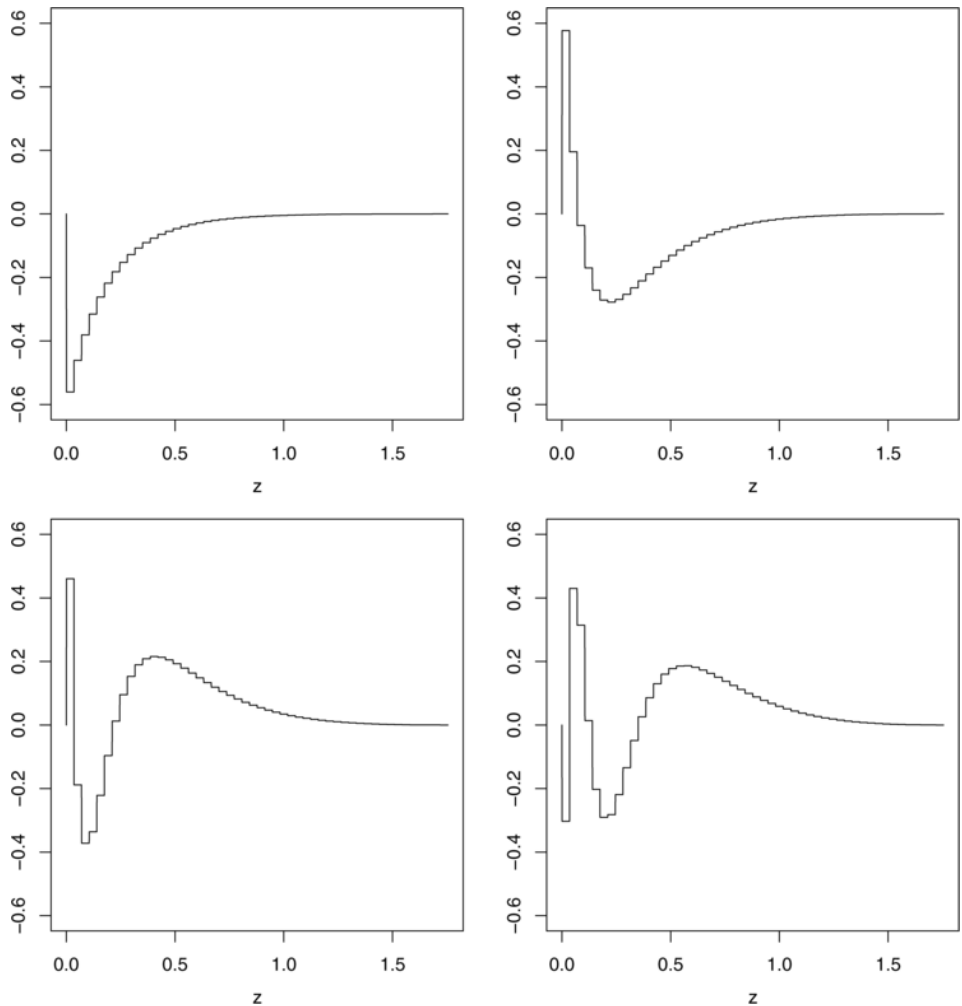

FIG. 3. *The first four Fisher basis functions, all of which decay quickly to zero in redshift.*

It was for this reason that we defined the $\mathcal{M}_k$ as above. Each $\mathcal{M}_k$ contains constant and smooth functions over the entire redshift domain. As a consequence, a good estimate of a smooth function can be obtained in any of the models, and when $w$ is in fact smooth, the procedure will select a lower-dimensional model, with consequent gains in mean squared error. Moreover, unlike polynomials, the B-splines are localized and numerically well conditioned.

Of course, for most purposes, an estimate of $w$ is not sufficient, so we also need an assessment of uncertainty in the estimates. Consider first inference for a single $\mathcal{M}_k$. It is possible to use a Normal approximation at the maximimum likelihood estimate from the nonlinear regression to derive an approximate covariance matrix for the estimated parameters. In practice, this seems to perform well, but because it is difficult to bound the accuracy of the approximation given the nonlinearity, we use a resampling approach. The error bars can be computed using a parametric or nonparametric bootstrap. We prefer to use a nonparametric bootstrap since the error bars will then be less affected by any bias in the specification of the model, especially given that we are selecting among models initially. We generate bootstrap confidence intervals for the parameters in the model by resampling residuals from the model fit, renormalized to have appropriate variance [Efron (1979), Efron and Tibshirani (1994)]. The basic procedure is as follows:

1. Compute the maximum likelihood estimator $(\hat{\beta}, \hat{\theta})$, where $\beta$ is the vector of parameters for $w$ and $\theta = (H_0, \Omega_m)$.
2. Compute residuals $e_i = Y_i - \hat{r}(z_i; \hat{\beta}, \hat{\theta})$.
3. Using the standard errors of $Y_i$ and the linear approximation at the maximum likelihood estimator, standardize the residuals to unit variance. Call these standardized residuals $\varepsilon_i$.
4. For $b = 1, \ldots, B$, for some large $B$, draw pseudo-noise from the empirical distribution of the $\varepsilon_i$. Call these $\varepsilon_i^{*(b)}$ for $i = 1, \ldots, n$.
5. Generate pseudo-data

$$
(41) \qquad Y_i^{*(b)} = \hat{r}(z_i) + \sigma(z_i)\varepsilon_i^{*(b)}.
$$

6. Compute the maximum likelihood estimates $(\hat{\beta}^{*(b)}, \hat{\theta}^{*(b)})$ from each pseudo-data set.
7. Compute standard errors and confidence intervals for these parameters from the $(\hat{\beta}^{*(b)}, \hat{\theta}^{*(b)})$'s as in Efron (1979) and Efron and Tibshirani (1994).

We use the bootstrap confidence intervals to compute confidence bands for $w$ and $q$ by computing the largest and smallest values of the functions at each redshift that are consistent with the confidence intervals on the parameters.

In the nonparametric case, it is common practice to use the confidence bands corresponding to the selected model. These are straightforward and accurate when the selected model holds but are necessarily optimistic because the bands do not account for the variation in the model selection process or for the potential bias induced by choosing too simple a model.

A simple improvement we use is to incorporate the model selection into the resampling process, using the largest and smallest estimated $w$ or $q$ from the bootstrap samples to construct the confidence bands. Because this effectively includes model bias in the bands, this approach is likely to be somewhat conservative. We will explore other approaches to this problem in a future paper.

## 7. Results.

7.1. *Testing cosmological models.* We test seven cosmological models using the procedure described earlier, independently of any parameterization for $w$. Three models (cosmological constant, frustrated cosmic strings, and domain walls) can be tested with a simple null hypothesis of the form $w = w_0$. The two quintessence models (thawing and freezing solutions, resp.) were tested with inequality null hypotheses given by equations (36) and (37) intersected with the condition that $-1 \leq w \leq -0.8$. [We also tested more expansive versions of these hypotheses using (i) equations (36) and (37) alone and (ii) the hypothesis $w' \geq 0$ and $w' \leq 0$ intersected with the condition that $-1 \leq w \leq -0.8$. But being strict supersets of the original null hypotheses, these are less likely to reject.] We tested both the matter dominated and strongly nonaccelerating universe hypotheses. For the latter, we used a confidence interval obtained from the current best estimates[13] and adjusted the confidence level as described in Section 5. Finally, we tested the inclusion hypothesis that $w$ is a constant, possibly different from $-1$. Table 1 shows the results of these tests at various significance levels. The no dark energy model is clearly inconsistent with the data ($p$-value $p \approx 0$), but none of the other models are rejected at the 13% level. Note, in particular, that the cosmological constant is consistent with the data.

A false null hypotheses might fail to be rejected because the power of the test is too low. Because our procedure has essentially as much power as possible given the available information about $w$, the only ways to improve power are either to make stronger assumptions about the form of $w$ or to get more data. We argue that the latter is necessary. The results do not change when performing the same tests assuming a linear form $w(z) = -(\beta_0 + \beta_1 z)$, which is the simplest nontrivial parameterization and a correspondingly smaller confidence set. The pattern of rejections is basically the same, and in particular, there is insufficient evidence to move away from a cosmological constant. Of course, there is no reason to believe the linear form for $w$, and if it is false, inferences under that assumption can be misleading. But this shows that strengthening the assumptions is not enough to overcome the lack of information in the data.

---

[13]For current estimates of cosmological parameters, see http://lambda.gsfc.nasa.gov/product/map/current/. We adopt the value of $H_0^2 \Omega_m$ derived from "all" data for the LCDM model. This gives $H_0$ $72 \pm 8$ km/s/Mpc. Note, however, that the absolute value of the Hubble constant is only determined by the supernova data up to an arbitrary shift because of calibration of the absolute magnitudes. We thus recentered this confidence interval around 65, which is consistent with the calibration of our data.

TABLE 1
*Results of nonparametric hypothesis tests for various cosmological models of $w$. The significance levels correspond to 1, 1.5, 2, and 2.5 standard deviations respectively from a Gaussian mean*

| Model | Rejected at level | | | |
| --- | --- | --- | --- | --- |
| | **32%** | **13%** | **5%** | **1%** |
| Cosmological constant | **yes** | no | no | no |
| Frustrated cosmic strings | **yes** | yes | yes | no |
| Domain walls | **yes** | no | no | no |
| Matter dominated | **yes** | yes | yes | yes |
| Nonaccelerating | **yes** | yes | yes | no |
| Quintessence thawing | no | no | no | no |
| Quintessence freezing | no | no | no | no |
| Constant $w$ | no | no | no | no |

7.2. *Fitted models for the equation of state.* Our fitted $w$ is a constant, $\hat{w} \equiv -1.013 \pm 0.124$ with $\hat{\Omega}_0 = 0.268 \pm 0.028$ and $\hat{H}_0 = 65.6 \pm 0.90$, where the standard errors are based on 1000 bootstrap iterations. (Note that the supernova data determine the value of the Hubble constant plus an arbitrary shift induced by calibration of the supernova absolute magnitudes, so it is the relative uncertainty rather than the absolute value that matters here.) The bootstrap $BC_a$ 95% confidence intervals do not differ much from the Normal intervals based on the bootstrap standard errors: $[-1.262, -0.796]$ for $w$, $[0.220, 0.324]$ for $\Omega_m$, and $[63.9, 67.4]$ for $H_0$. It is straightforward to compute joint confidence sets for these parameters, but we do not report them here. As a check on these results, we note that within the parametric models we consider (polynomials in $z$ or $a$, piecewise constants, B-splines) likelihood ratio tests between the constant model and the higher-order models in the family fails to reject with $p$-value $p > 0.85$. In all three cases, BIC is monotone increasing with the constant model the clear choice. A likelihood ratio test of the cosmological constant versus the constant $w$ model fails to reject with $p$-value $p = 0.137$.

7.3. *The need for more data.* A key question is whether current supernova data are sufficient to resolve the differences among interesting models for $w$ and $q$. We argue here that the answer is no.

TABLE 2
*Power for distinguishing a constant model $w \equiv w_1$ from a cosmological constant $w \equiv -1$ using a likelihood ratio test with significance level $\alpha$. Each is based on 3000 simulations of data from equation (20)*

| Alternative ($w_1$) | Significance level ($\alpha$) | | | |
|---|---|---|---|---|
|  | 0.68 | 0.87 | 0.95 | 0.99 |
| $-0.85$ | 0.352 | 0.169 | 0.080 | 0.015 |
| $-0.90$ | 0.328 | 0.157 | 0.056 | 0.012 |
| $-0.95$ | 0.303 | 0.121 | 0.047 | 0.012 |
| $-0.99$ | 0.322 | 0.116 | 0.044 | 0.007 |

First, even under strong assumptions and with essentially optimal procedures, there is not enough evidence to distinguish among interesting models. The cosmological constant model is suggestively on the boundary at the 13% level, but no conclusive differences are supported by the data.

Second, we can use as a minimal criterion for resolvability the power of the likelihood ratio test for distinguishing the cosmological constant model from a constant $w$ model. With existing standard errors, it is straightforward to compute this power by simulation through equations (21) and (20). Table 2 shows the power of this test for various significance levels and alternatives, all of which are low. The power for distinguishing a constant $w$ model from a piecewise constant model with one breakpoint are lower for a similar variety of alternatives.

Third, an even more striking demonstration of model degeneracy is given by Figure 4. This shows two very different equations of state that give virtually indistinguishable fits to the data, with a chi-squared deviation of 0.04. This example is driven primarily by uncertainty in $\Omega_m$, which can be reduced using other (non SNe) data.

**8. Discussion.** Matching the many studies in the astrophysics and cosmology literature, we find that current supernova data do not yet provide tight enough inferences to make nontrivial claims about $w$. And in particular, the data are consistent with a cosmological constant model. This puts the focus squarely on future data, which will likely provide notably stronger constraints on the dark energy equation of state. This raises the question of what this newfound precision will mean. If the cosmological constant model holds or if $w$ is very close to but not equal to $-1$, then we will be in much the same state as we are currently. But otherwise, the insight we gain into the nature of dark energy will depend on our ability to distinguish competing models and to infer subtle structure in $w$. For this purpose, a nonparametric approach will be particularly effective. This paper describes a new technique for testing among competing dark energy models with minimal assumptions and describes a framework for nonparametric estimation of $w$.

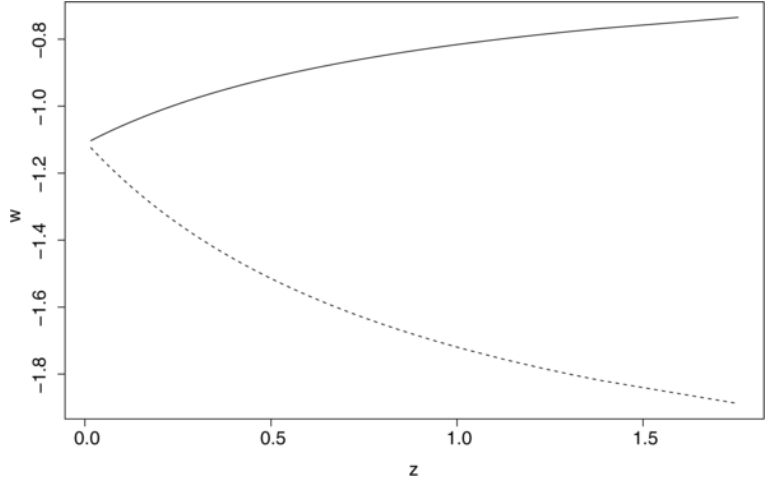

Fig. 4. *Two equations of state that produce statistically indistinguishable r's.*

Several challenges and open questions remain. Computation of sharp, honest confidence sets in nonlinear inverse problems is a mathematically difficult problem but important to getting as much information as possible from the data. Expanding the scope of the hypothesis testing methods to new classes of hypotheses that eliminate "unphysical" possibilities will improve the power of the technique. And systematic errors will become a major issue as the statistical uncertainties become smaller.

The next generation of supernova data sets may answer many questions with the methods presented herein, but they will also raise a host of interesting new statistical problems.

## APPENDIX: EQUATION OF STATE BOUNDS FOR THE QUINTESSENCE MODELS

To derive equations (36) and (37), we begin by transforming equations (34) and (35) from the scale factor $a$ to redshift $z$. Replacing $dw/d\ln a$ by $-(1+z)w'(z)$ and reversing the inequality because of the negative sign yields the corresponding equations

$$(42) \quad -3\frac{1+w(z)}{1+z} \leq w'(z) \leq -\frac{1+w(z)}{1+z},$$

for thawing solutions, and

$$(43) \quad -0.2w(z)\frac{1+w(z)}{1+z} \leq w'(z) \leq -3w(z)\frac{1+w(z)}{1+z},$$

for freezing solutions.

Begin with the assumption that $w > -1$, which we will weaken below. For the thawing equalities, divide through by $1 + w$ to get $w'/(1 + w) = (\log(1 + w))'$ and, thus,

$$(44) \qquad -3\frac{1}{1+z} \leq (\log(1+w(z)))' \leq -\frac{1}{1+z}.$$

Integrating through from 0 to $z$ yields

$$(45) \qquad -3\log(1+z) \leq \log(1+w(z)) - \log(1+w(0)) \leq -\log(1+z),$$

and taking exponents,

$$(46) \qquad (1+z)^{-3} \leq \frac{1+w(z)}{1+w(0)} \leq (1+z)^{-1},$$

which leads directly to equation (36).

Similarly for the freezing solutions, $w'/w(1+w) = (\log(w/(1+w)))'$. Dividing through and integrating as before gives

$$(47) \quad -0.2\log(1+z) \leq \log\left(\frac{w(z)}{1+w(z)}\right) - \log\left(\frac{w(0)}{1+w(0)}\right) \leq -3\log(1+z).$$

Taking exponents and simplifying gives equation (37).

For freezing solutions, $w' \geq 0$, and if $w(0) = -1$, then $w \equiv -1$. For thawing solutions, either $w(0) = -1$ and $w \equiv -1$, or $w(z) = -1$ for some $z > 0$. The latter case leads to a contradiction given the bounds on $w'$ and continuity of $w$. Hence, the bounds hold for $-1 \leq w \leq 0.8$.

**Acknowledgment.** The authors would like to thank the referee for an extremely thorough and helpful review and many good suggestions. The paper is much improved as a result.

## SUPPLEMENTARY MATERIAL

**On-line supplementary material for Genovese et al. 2009** (DOI: 10.1214/08-AOAS229SUPP; .pdf). We provide further technical details on data issues, derivations, and methods.

C. R. Genovese
P. Freeman
L. Wasserman
Department of Statistics
Carnegie Mellon University
USA
E-mail: genovese@stat.cmu.edu
E-mail: pfreeman@stat.cmu.edu
E-mail: larry@stat.cmu.edu

R. C. Nichol
Institute of Cosmology and Gravitation
University of Portsmouth
England
E-mail: bob.nichol@port.ac.uk

C. Miller
CTI/NOAO
950 North Cherry Avenue
Tucson, Arizona 85719
USA
E-mail: cmiller@noao.edu